\documentclass[
    preprint,
    3p, %3+ style journals
    twocolumn,
    sort&compress,
]{elsarticle}
\usepackage[utf8]{inputenc}
\usepackage{lipsum}
\usepackage[colorlinks,linkcolor=blue,citecolor=blue,urlcolor=blue]{hyperref}
\usepackage{amsmath}
\usepackage{amsfonts}
\usepackage{amssymb}
\usepackage{hyperref}
\usepackage{cleveref}
\usepackage{graphicx}
\usepackage{mhchem}
\usepackage{cuted}
\usepackage[toc,page]{appendix}
\makeatletter
\def\@author#1{\g@addto@macro\elsauthors{\normalsize%
    \def\baselinestretch{1}%
    \upshape\authorsep#1\unskip\textsuperscript{%
      \ifx\@fnmark\@empty\else\unskip\sep\@fnmark\let\sep=,\fi
      \ifx\@corref\@empty\else\unskip\sep\@corref\let\sep=,\fi
      }%
    \def\authorsep{\unskip,\space}%
    \global\let\@fnmark\@empty
    \global\let\@corref\@empty  %% Added
    \global\let\sep\@empty}%
    \@eadauthor={#1}
}
\makeatother

\title{Reduction of detection limit and quantification uncertainty due to interferent by neural classification with abstention}
\author[pnnl]{Alex Hagen\corref{cor1}}
\cortext[cor1]{alexander.hagen@pnnl.gov}
\author[pnnl]{Ken Jarman}
\author[pnnl]{Jesse Ward}
\author[pnnl]{Greg Eiden}
\author[pnnl]{Charles Barinaga}
\author[pnnl]{Emily Mace}
\author[pnnl]{Craig Aalseth}
\author[pnnl]{Anthony Carado}
\address[pnnl]{Pacific Northwest National Laboratory, Richland, WA, USA}
\date{\today{}}

\begin{document}

\begin{abstract}
Many measurements in the physical sciences can be cast as counting experiments, where the number of occurrences of a physical phenomenon informs the prevalence of the phenomenon's source.  Often, detection of the physical phenomenon (termed signal) is difficult to distinguish from naturally occurring phenomena (termed background).  In this case, the discrimination of signal events from background can be performed using classifiers, and they may range from simple, threshold-based classifiers to sophisticated neural networks.  These classifiers are often trained and validated to obtain optimal accuracy, however we show that the optimal accuracy classifier does not generally coincide with a classifier that provides the lowest detection limit, nor the lowest quantification uncertainty.  We present a derivation of the detection limit and quantification uncertainty in the classifier-based counting experiment case. We also present a novel abstention mechanism to minimize the detection limit or quantification uncertainty \emph{a posteriori}. We illustrate the method on two data sets from the physical sciences, discriminating Ar-37 and Ar-39 radioactive decay from non-radioactive events in a gas proportional counter, and discriminating neutrons from photons in an inorganic scintillator and report results therefrom.
\end{abstract}

\maketitle

\section{Motivation}

Many physical measurements consist of counting experiments (CEs), where the rate of occurrence of an event informs quantitative information about a physical system.  These experiments are performed by discretely counting these events over a designated counting time. The two main goals of such CEs are to either detect the presence of a given phenomenon in a physical system, or to measure the prevalence of a phenomenon; the performance of tasks which are best indicated by the detection limit and measurement uncertainty, respectively. Counting techniques underlie many physical sciences. Examples include the measurement of mass or specific activity of an isotope in a mixed sample, the measurement of the ratio of neutron to photon doses in radiological experiments, or even the prevalence of a given pathogen in a population of people.

All but the most trivial counting experiments suffer from the presence of "background" events, where an unrelated event is detected and thus counted along side the "signal" events of interest. The separation of background from signal can be effected in many ways. The three main classes are: physical removal of interference, such as movement of a radiation counting experiment into underground labs to occlude cosmic rays; experiment design, such as differential sensitivity measurements which identify differing sensitivity changes to analyte versus interferent; and data analytic methods, such as the removal of events exhibiting the characteristics of radon in radioxenon measurements \cite{Hagen2021}. In many cases, however, physical removal of interference or careful experiment design to remove background is either impossible (as is the case in counting experiments in the dark matter search), or cost and effort prohibitive (as is the case in many material separation studies). As machine learning (ML) techniques have matured, the data analytic methods for signal and background separation have become increasingly sophisticated and exhibited higher performance \cite{Mace2018, Hagen2021, Parsons2020, Pearkes2017, Renner2017}. These methods are also receiving heightened interest as a cost-saving and throughput-increasing measure.

An important class of data analytic methods is that of event-by-event classification into signal or background classes, which we will refer to as Classifier-Based Counting Experiments (CBCE).  This structure nicely aligns the problem in the counting statistics domain with a main category of problems in general data science - that of binary classification. Many advances in binary classification are readily applicable to CBCEs.  Unfortunately, in the binary classification literature, classifiers are rarely perfect, and the best are often chosen as those that have the highest accuracy amongst all tested methods\footnote{In this context, accuracy has the narrow definition of simply the ratio between the correctly classified events and all events.}.  We show that, for a non-perfect classifier, maximal accuracy is not optimal for minimizing quantification uncertainty. We also reprise \cite{Hagen2021}, showing that maximal accuracy does not minimize the detection limit. As detection limit and quantification uncertainty are the most important measures of performance in counting experiments, we seek to remedy these findings.

We contribute a novel method for improving both the quantification uncertainty and detection limit of any classifier by developing an optimal threshold, based on the classifier's raw output, which maximizes the metric of interest. This method, which we call abstention, can be applied to any CBCE with modest assumptions about the signal-to-background ratio (SBR). While it is beneficial at any SBR, it can realize orders of magnitude improvement in the quantification uncertainty in the case of very low SBR.

To illustrate the use of abstention in CBCEs, we structure this work in the following way.  We identify related work and this work's antecedents.  Then, we illustrate \emph{non-optimality of detection limit using a maximal-accuracy classifier} on a data set consisting of events induced by neutrons and photons in a scintillator.  We follow that with the process for determining the thresholds(s) obtaining the minimal uncertainty for a classifier operating on events from an ultra-low background proportional counter measurement.  To reinforce our claim that abstention can be used on any classifier with a continuous output, we use different classifiers throughout. For narrative purposes, details of these classifiers are kept to a minimum.

\section{Relevant Literature}

A large uncertainty quantification literature, and in fact a sub-field of statistics, exists to determine the proper way to quantify samples in the presence of background. The statistical methods for quantifying uncertainty and determining detection limits given signal and background properties are mature and well founded, progressing to the point of technical manuals describing best practices \cite{MARLAP}.  This, however, is not true of the methods for CBCE.  Statistical approaches to the uncertainty propagation through a CBCE scheme are sparse in the literature, and the literature includes even fewer attempts to improve quantification uncertainty in a CBCE.

Our method for optimizing the quantification uncertainty depends on abstention from certain data points based on the classifier output (more strictly the classifier's confidence in its own output), and some assumptions about the system in question.  Similar forms of abstention have existed in the machine learning literature since the 1970s. Hellman derived a special form of Nearest Neighbors which used several methods to abstain from classification \cite{Hellman1970}. More recently, abstention has dealt with increasing the robustness of a classifier to mistakes within the labels of a data set, while continuing to measure classifier performance with accuracy.  Thulasidasan provides several methods for this, including one which modifies the training of a neural network for this robustness \cite{Thulasidasan2019, Thulasidasan2019a}. These works developed the basis for our work: the ability of a ML based classifier to abstain from classifying certain data.  However, while the previous efforts focused on developing abstention for specific ML classifiers, and rely on accuracy for their performance metrics, we instead focus on tying a simple method for abstaining from classification to two important physical metrics.

We develop further on a line of investigation started by Chow in 1970. Chow described the reject rate and error tradeoff for an ideal optical character recognition system \cite{Chow1970}. We use this tradeoff concept throughout the present work.  Recently, DeStefano generally described that the metric for any classifier must necessarily be dependent on the rejection or abstention threshold \cite{DeStefano2000}, which Geifman applied to neural networks \cite{Geifman2017}.  We modify these works in two ways, applying them specifically to the counting experiment domain, while generalizing to any classifier with a continuous output (which we call "scores").

\section{Detection Limit}\label{sec:detection_limit}

A common use for counting experiments is to determine the presence of a given phenomenon; a use case which pervades many fields, including nuclear forensics, beyond-standard-model physics, and even medical diagnostic applications.  For these cases, the detection limit, or minimum detectable amount, is the metric of interest for a given detection methodology.  We show below that maximum accuracy thresholds on CBCEs are in general not coincident with optimal detection limit thresholds, and illustrate this fact on an example data set.

\subsection{Illustrative Example: Neutron versus photon discrimination in scintillators}

\paragraph{Motivation} The efficient detection of neutrons is an important subfield of nuclear safety, special nuclear material interdiction, and radiotherapy. Often, however, the neutron detection mechanism coincides with detection mechanisms for photons; most neutron detectors thus have a large "photon background".  One common way to solve this issue is to use scintillators, which emit light with differing properties when a neutron or photon interacts with the material.  Then, the resulting light distribution is digitized, and so-called "Pulse Shape Discrimination" (PSD) methods are employed \cite{knoll2010radiation}. The classification of an event as originating from a neutron or a photon is a challenge and the performance possible is dependent on the detection material's response to each species, the temperature and other environmental factors of the experiment, and, as we subsequently show, the performance of a classifier.  We show that, while creating a classifier for PSD is straightforward, the optimization of the threshold above which that classifier classifies an event as neutron can have significant effects on the detection limit.

\begin{figure}
    \centering
    \includegraphics{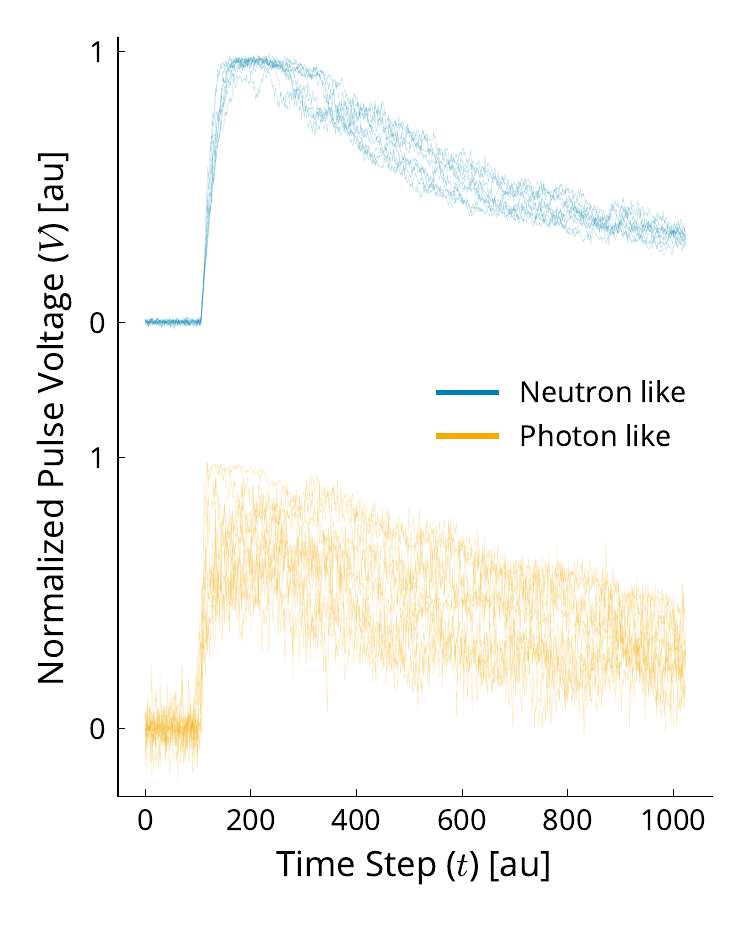}
    \caption{Example pulses emanating from interaction of a neutron or photon with CLYC scintillator material}
    \label{fig:n_vs_g_pulses}
\end{figure}

\paragraph{Data} We obtained a data set of digitized pulses collected from a $\ce{Cs}_{2}\ce{Li}\ce{Y}\ce{C}_{6}$ (CLYC) detector when exposed to mostly photon and mostly neutron sources.  An example of these pulses is shown in \cref{fig:n_vs_g_pulses}.  The relevant feature to these pulses is the curvature around the maximum magnitude of the pulse: a pulse with a fast rise time and flat area at the top of the curve is indicative of a neutron event, an event with no flat area at the top of the curve is indicative of a photon event.  This difference is physically based - a neutron interacts with CLYC via the $\ce{^{6}Li} + n \rightarrow \alpha + \ce{^{3}H}$ reaction. The $\alpha$ product deposits most of its energy within nanometers of the original event, which in turn creates a slowly decaying burst of scintillation photons; photons interact with CLYC by liberating electrons, which decay away faster, creating a more quickly decaying burst of scintillation photons. 

\begin{figure}
    \centering
    \includegraphics{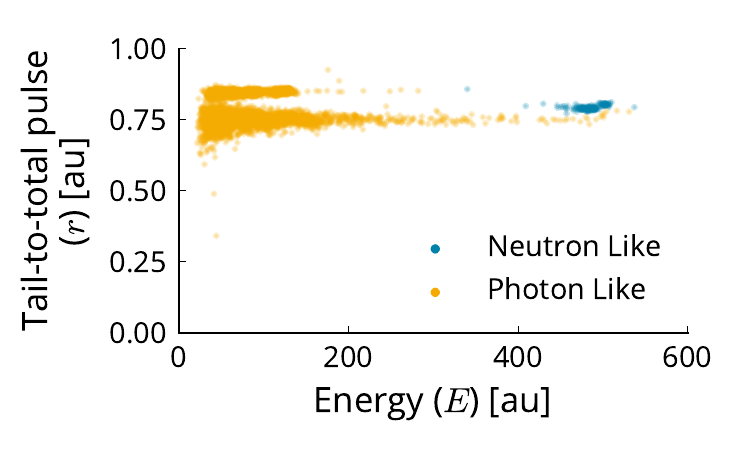}
    \caption{Tail-vs-total versus total pulse for neutron and photon like pulses. One can see that in general, a neutron like pulse has a lower tail-over-total value than those from photons.}
    \label{fig:tail_over_total}
\end{figure}

\paragraph{Classification} Discrimination of neutrons and photons is complicated by the near impossibility of generating a pure-neutron source: practically, any experiment will provide detection pulses of mixed neutron and photon origin.  We resolve this ambiguity by using a dimensionality reduction method for identifying neutron and photon sources within the data set, treating these as our ground-truth labels\footnote{Note that, while we strive for physical accuracy in our labels, the following derivation applies equally to any set of labels; therefore exact physical accuracy is a secondary concern for labeling in the current work.  We provide details about our method in Appendix~\ref{sec:totc_labeling}}.  We identify neutrons as "analyte" and photons as "interferent".  We perform standard transformations to each pulse, to calculate the "total" pulse (integrating) and the "tail-over-total" ratio (integrating the tail of the pulse and dividing by the total pulse) as follows:
\begin{equation}\label{eq:total_pulse}
    P = \int_{0}^{t_{p}} V\left(t\right)dt,\quad D = \int_{t_{p}}^{t_{d}} V\left(t\right)dt
\end{equation}
\begin{equation}\label{eq:tail_to_total_pulse}
    r = \frac{D}{P + D}
\end{equation}
where $V\left(t\right)$ is the measured voltage of each pulse at time $t$, $t_{p}$ is the end time of some "prompt" window in the pulse, and $t_{d}$ is ending time of some "delayed" window in the pulse.  In practice, these integrals are approximated as simple summations. We compare this to the energy of the event, which is simply the maximum value of the pulse above its baseline.
\begin{equation}
    E = \mathrm{max}\left[V\left(t\right)\right]
\end{equation}
These features are visualized in \cref{fig:tail_over_total}. 

\begin{figure}
    \centering
    \includegraphics{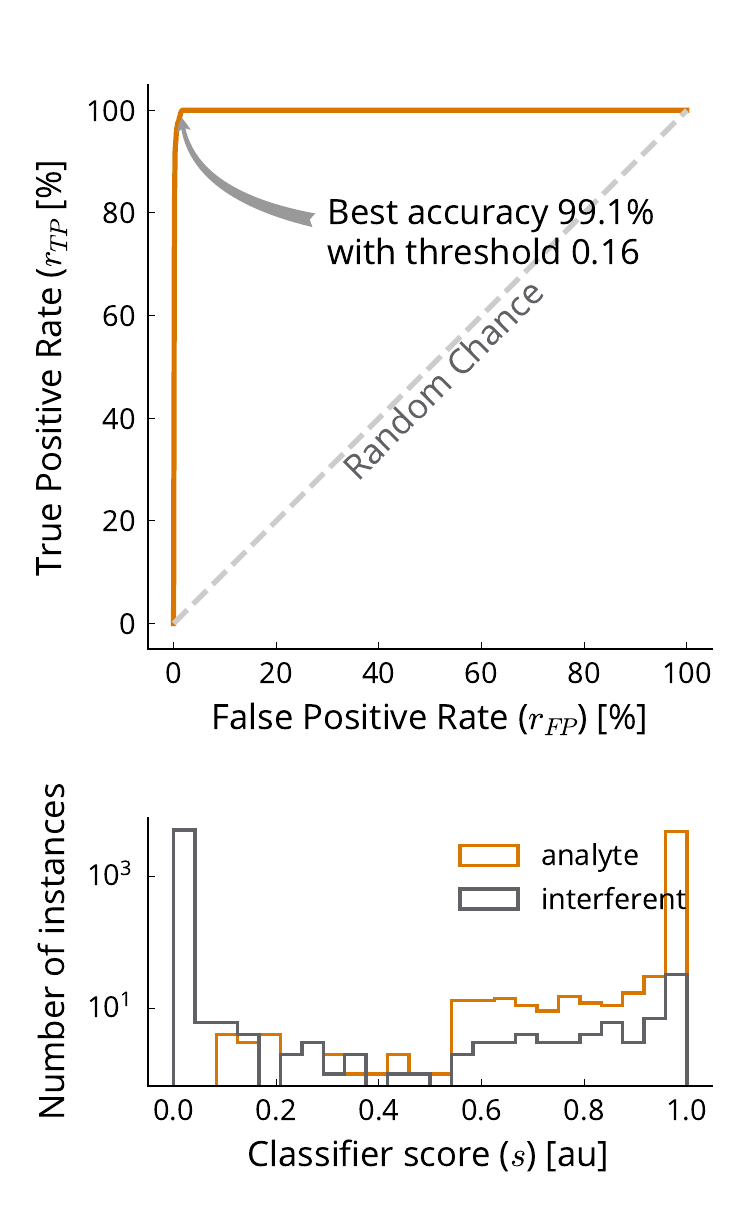}
    \caption{Receiver operating characteristic for TOTC (top) and associated histograms (bottom). The classifier performs well when measured by accuracy or area under the ROC curve. The scores attributed to neutron versus photon events are separated but not disjoint.}
    \label{fig:totc_roc}
\end{figure}

We then create a classifier, called the tail-over-total classifier (TOTC). The TOTC uses the two features shown on \cref{fig:tail_over_total} as input and attempts to classify whether an event is neutron-like or photon-like.  TOTC is a multilayer perceptron, and performs well when measured by the area under the receiver operating characteristic curve.  The receiver operating characteristic (ROC) curve is shown in \cref{fig:totc_roc}, alongside a histogram of classifier scores for analyte and interferent.

The second panel of \cref{fig:totc_roc} shows a characteristic typical of counting experiments.  While the distribution of scores for analyte and interferent are quite different, they are not disjoint. This characteristic leads directly to the main thesis of this paper: that maximum accuracy is not a good metric for CBCEs, and by consequence the threshold for classification should not optimize accuracy.  To show this, we illustrate that the detection limit of a neutron source given TOTC is not minimized by the maximal-accuracy threshold.

\paragraph{Detection Limit Optimization}

We calculate the detection limit following \cite{Hagen2021}, which draws from \cite{Currie1968} with some corrections (such as the Stapleton correction \cite{MARLAP}) for low signal counts.  Then, the detection limit $l_{d}$ is given by
\begin{equation}\label{eq:mda}
    l_{d} = \frac{l_{c} + \frac{k^{2}}{2} + k\sqrt{\frac{k^{2}}{4} + l_{c}}}{\eta R T}
\end{equation}
where $k$ is Z-score for a false alarm rate of 5\%, $R$ is the expected rate of interferent in a counting experiment of time $T$, $\eta$ is the efficiency of detecting analyte given a threshold, and $l_{c}$ is the critical limit calculated by
\begin{equation}\label{eq:lc}
    l_{c} = 2.33\sqrt{n_{IA} + 0.4} + 1.35
\end{equation}
where $n_{IA}$ is the number of interferent detections classified as analyte for a given threshold, and $0.4$ is the Stapleton correction for low analyte detection rates.

\begin{figure}
    \centering
    \includegraphics{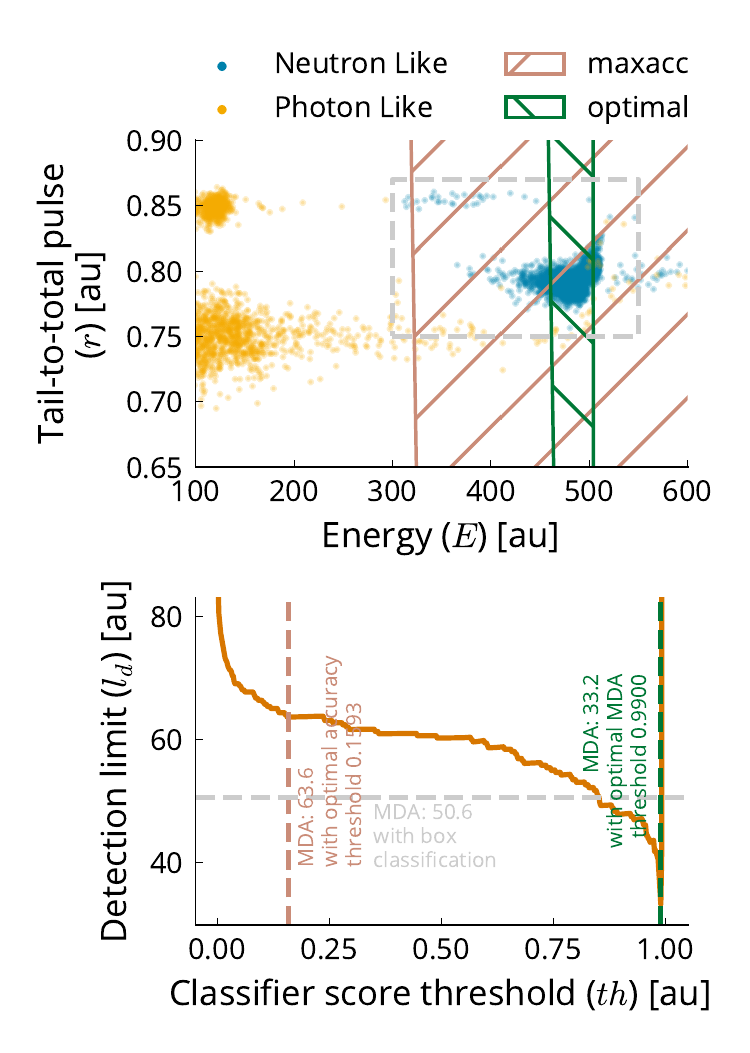}
    \caption{Detection limit versus classifier score low limit when the interferent is $10,000\times$ more prevalent than the analyte. A classification created by visual inspection (dubbed "box" classification", shown in grey on the top panel) results in a detection limit close to twice that of the optimal detection limit (shown in green on the top panel). The maximum accuracy threshold (shown in red on the top panel) for classification as analyte results in a similar detection limit to that of the box classification. }
    \label{fig:totc_mda}
\end{figure}

We can then profile the detection limit versus a varying classifier threshold by calculating $l_{d}$ for varying thresholds.  The results of this calculation are presented in \cref{fig:totc_mda}. This figure illustrates several expected properties of the detection limit. When the threshold is too high, very few pulses are classified as analyte, and a very high detection limit results.  When the threshold is too low, a high detection limit also results because of a high rate of classification interferent pulses as analyte.

Regarding the hypothesis of this paper, \cref{fig:totc_mda} explicitly shows that the maximal-accuracy threshold is not optimal regarding detection limit. In fact, for the example provided, the detection limit for the maximal-accuracy threshold is twice as high as the optimal limit.

\section{Measurement Uncertainty}

Another large class of counting experiments are those used to quantify a material or phenomenon of interest, beyond simply detecting it.  This technique is again used in a broad variety of fields, from nuclear forensics to prediction of political election results.

The results of such counting experiments, if analyzed without an estimate of interferent prevalence, can lead to extremely biased results.  For example, a classifier discriminating analyte from interferent which has 90\% accuracy, when applied to a population of 100,000 members, 99\% of which are interferent, will on average identify 10,000 members as analyte, 9,000 of which are actually interferent.  This is exactly the base-rate fallacy \cite{Kahneman1973}, and utilization of statistical corrections for this fallacy are essential to accurate quantification.

CBCEs are one broad class of correction for counting experiments with irreducible interference.  The use of \emph{a priori} measurements of the performance of the classifier involved lead to a more accurate estimate of the analyte prevalence.  In order to correct for classifier performance, we can use the method of moments. We first calculate the proportion of correct prediction of analyte ($p_{a}$) and interferent ($p_{i}$) and the proportion of incorrect prediction of analyte ($q_{a}$) and interferent ($q_{i}$).  Then, we can state the expected value of counts classified as analyte and interferent as
\begin{equation}\label{eq:Ya}
    Y_{a} = p_{a}\mu_{a} + q_{i}\mu_{i}
\end{equation}
\begin{equation}\label{eq:Yi}
    Y_{i} = q_{a}\mu_{a} + q_{i}\mu_{i}
\end{equation}
where $\mu_{a,i}$ are the prevalence of analyte and interferent, respectively.  Then, this reduces to
\begin{equation}\label{eq:mua}
    \mu_{a} = \frac{p_{a} Y_{a} - q_{i} Y_{i}}{p_{a}p_{i}-q_{a}q_{i}}
\end{equation}

The uncertainty of this unbiased estimator is then derived using the conditional variance formula \cite{Ross2014}
\begin{equation}
    \mathrm{Var}\left(\mu_{a}\right) = \mathbb{E}\left[\mathrm{Var}\left(\mu_{a} | X_{a}, X_{i}\right)\right] + \mathrm{Var}\left(\mathbb{E}\left[\mu_{a}|X_{a}, X_{i}\right]\right)
\end{equation}
where $\mu_{a}$ is the mean of counts from analyte, $X_{a}$ is a random variate denoting the counts from analyte, $X_{i}$ is a random variate denoting the counts from interferent, and $\mathbb{E}\left(\dots\right)$ denotes an expectation value.

By accounting for covariances (due to the constraint of sums on $X_{a}$ and $X_{i}$), considering counts to be multinomial random variables, and algebraic manipulation, we can derive the variance of $\mu_{a}$ as
\begin{multline}\label{eq:uncertainty_unbiased_mean}
    \mathrm{Var}\left(\mu_{a}\right) = \sigma_{a}^{2} + \frac{1}{\left(p_{a}p_{i} - q_{a}q_{i}\right)^{2}} \\
    \cdot \left\{ mu_{a}\left[p_{i}^{2}p_{a}\left(1-p_{a}\right) + q_{i}^{2}q_{a}\left(1-q_{a}\right) + 2p_{i}p_{a}q_{i}q_{a}\right]\right. \\
    \left. + \mu_{i}\left[p_{i}^{2}q_{i}\left(1-q_{i}\right)+q_{i}^{2}p_{i}\left(1-p_{i}\right) + 2p_{i}^{2}q_{i}^{2}\right]\right\}
\end{multline}
A full derivation of \cref{eq:uncertainty_unbiased_mean} is provided in Appendix~\ref{sec:derive_uncertainty}. For many counting experiments, $\mu_{a}$ can be assumed to be Poisson distributed, and in those cases $\sigma_{a}^{2}$ is equal to $\mu_{a}$. We use this approximation throughout. With that approximation, we reference the uncertainty as the standard deviation given that variance.

\begin{equation}\label{eq:uncertainty_definition}
    u \equiv \sqrt{\mathrm{Var}\left(\mu_{a}\right)}
\end{equation}

While it is not obvious whether the threshold for maximal accuracy minimizes \cref{eq:uncertainty_unbiased_mean}, we illustrate that it must not through an example from the physical sciences.

\subsection{Illustrative Example: Interferent rejection in Ultra Low Background Proportional Counters}

\paragraph{Motivation} A low background underground facility at Pacific Northwest National Laboratory has collected years of data measuring the prevalence of $\ce{^{37}Ar}$ and $\ce{^{39}Ar}$ for treaty verification and groundwater age dating, using ultra-low-background proportional counters (ULBPC) \cite{Aalseth2012}.  The isotopes of interest have very low prevalence, and difficult-to-reduce backgrounds such as micro-discharge events from signal paths and commercial electronics, along with natural background events from cosmic rays  can make it challenging to accurately quantify the activity of each isotope.  Mace, Ward, and Aalseth showed that a neural-network-based classifier could effectively separate the background noise events from the radioactive decay or gas-gain events \cite{Mace2018}.

\begin{figure}
    \centering
    \includegraphics{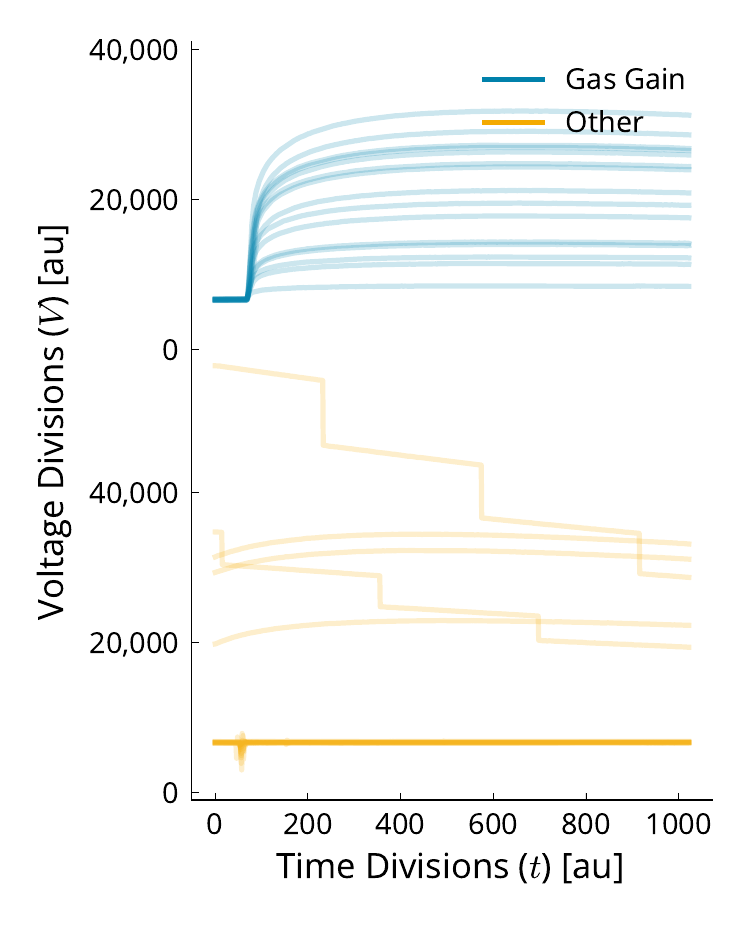}
    \caption{Example pulses from desired ("Gas Gain") and undesired ("Other") operation of the Ultra-Low Background Proportional Counter}
    \label{fig:ulbpc_pulses}
\end{figure}

\paragraph{Data} The ULBPC collects charge avalanches created in a gas under high electric field by energy depositions from nuclear decays, cosmic rays, and other sources.  These avalanches differ in total charge (energy) and timing (shape) due to their origin, with "Gas Gain" pulses starting from a baseline before a sharp rise and slow decay.  Other pulses, from various sources, have other shapes. Example pulse shapes for each origin are shown in \cref{fig:ulbpc_pulses}.

Data was also collected in two states where the detector was operated such that a higher number of bad pulses were collected than in the standard data set. We denote partitions of the standard data set as "Training" and "Validation" and the two states with higher bad pulse rate as "Alt. Mode 1" and "Alt. Mode 2". These alternate modes are used as test sets, to ensure that abstention not only reduces the uncertainty, but abstains from pulses in locations of low probability of correct classification in the general sense.

\paragraph{Methods}

In order to both discriminate Gas Gain from Other events, and to generalize well to events from the alternate modes of operation, we train a model using the Generalized ODIN method \cite{Hsu2020}. This architecture is a multilayer perceptron (details of which are provided in Appendix~\ref{sec:ulbpc}), which we train by minimizing cross-entropy between the known Gas-Gain/Other labels and the network predictions.

\begin{figure}
    \centering
    \includegraphics{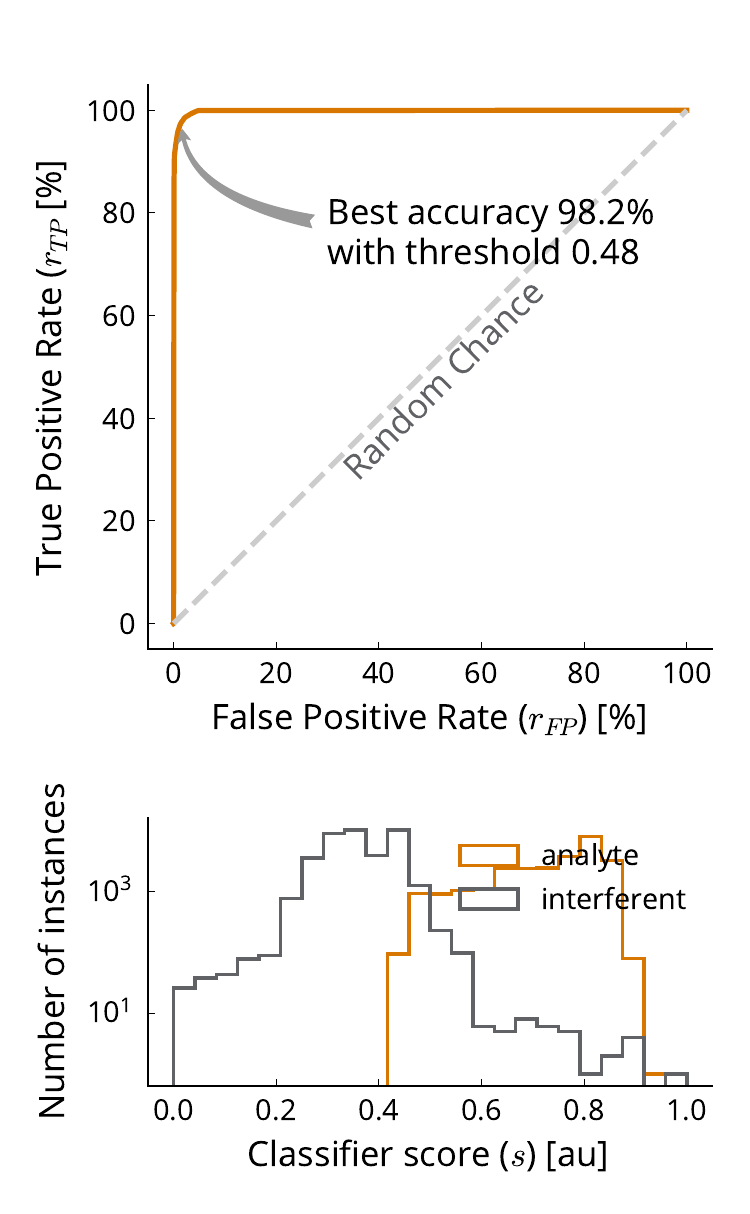}
    \caption{Performance of and classifier score distribution from ODIN network trained for ULBPC counting experiments. The network performs well when measured by area under the Receiver Operating Characteristic curve, with the analyte and interferent both having broad distributions of scores.}
    \label{fig:ulbd_roc}
\end{figure}

The resultant network again shows good performance when measured by area under the ROC curve or accuracy, and again shows separated but not disjoint score distributions for each class. This is shown in \cref{fig:ulbd_roc}.

\paragraph{Minimization of Uncertainty}

Given the trained ODIN classifier, we can then estimate the activity of the analyte, and the uncertainty thereof.  Calculation of the activity of the analyte is performed using \cref{eq:mua} and calculation of uncertainty is performed using \cref{eq:uncertainty_unbiased_mean}. Note that both of these equations require first the setting of thresholds $t_{a}$ and $t_{i}$ on classifier score.  Above $t_{a}$, all pulses are classified as analyte, and below $t_{i}$, all pulses are classified as interferent.  Then, using $p_{a}$, $p_{i}$, $q_{a}$ and $q_{i}$ calculated from a validation set, $\mu_{a}$ and $u$ can be estimated.

\begin{figure}
    \centering
    \includegraphics{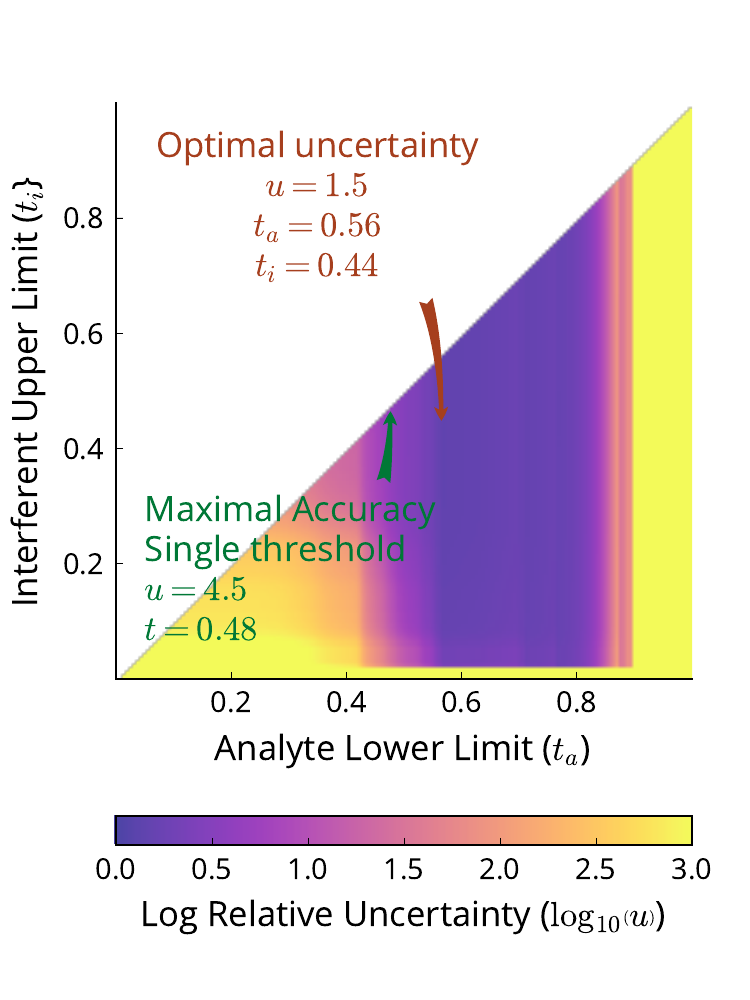}
    \caption{Uncertainty calculated given an analyte lower threshold of $t_{a}$ and interferent upper threshold of $t_{i}$ for analyte to interferent ratio of $10^{3}$.  The minimal value of uncertainty does not occur along the single threshold line (from $(0, 0)$ to $(1, 1)$), and is $3\times$ smaller than the uncertainty when using the maximal-accuracy threshold}
    \label{fig:ulbd_uncertainty_matrix}
\end{figure}

It is instructive first to examine how changing the thresholds affects the uncertainty for a given ratio of analyte to interferent $\frac{\mu_{a}}{\mu_{i}}$. We start with an examination of the case where there is $1000\times$ higher interferent activity than analyte. The uncertainty for this case is shown in \cref{fig:ulbd_uncertainty_matrix}, where the color indicates the base 10 logarithm of the uncertainty $u$, the location along the x-axis indicates $t_{a}$ and the location along the y-axis indicates $t_{i}$.  This shows that many threshold sets with low uncertainty do not fall along the single threshold region (the line from $(0, 0)$ to $(1, 1)$), and in fact the lowest uncertainty in this case is $\frac{1}{3}$ that of the uncertainty for the maximal-accuracy single threshold.

\begin{figure}
    \centering
    \includegraphics{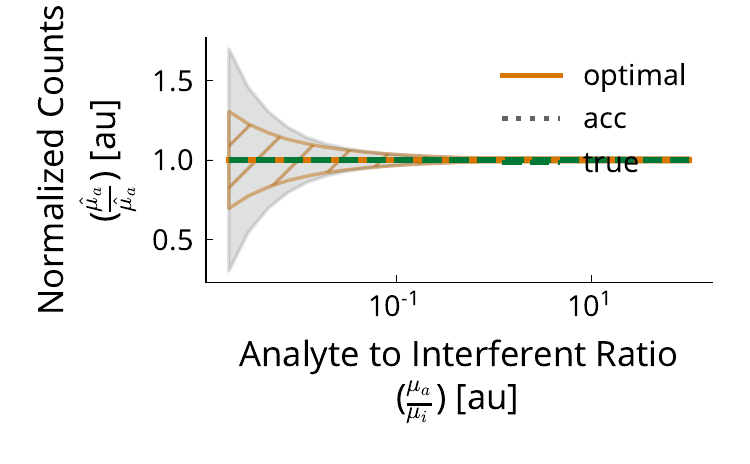}
    \caption{The estimated analyte activity per true analyte activity and its associated $1\sigma$ uncertainty when using the optimal and maximal-accuracy ("acc") threshold settings. The estimated activity of both methods does not differ from the true activity, but the uncertainty in the optimal case is much smaller than that in the maximal-accuracy case}
    \label{fig:ulbd_uncertainty_size}
\end{figure}

\begin{figure}
    \centering
    \includegraphics{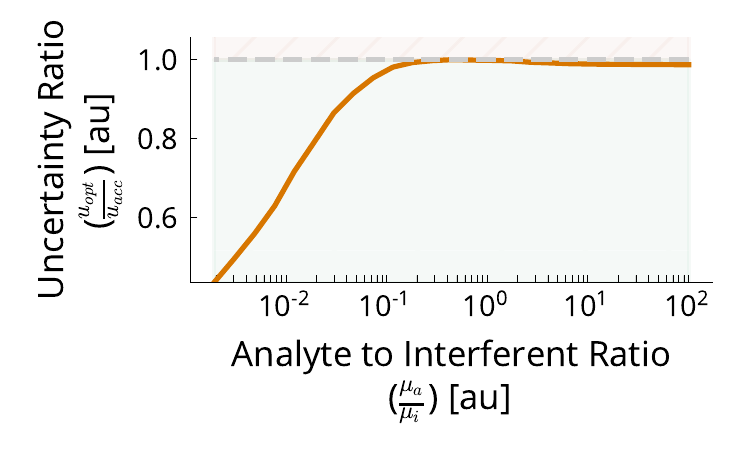}
    \caption{The ratio of the optimal to the maximal-accuracy uncertainty for varying analyte to interferent ratios.  At low analyte to interferent ratios, the optimal uncertainty is several times lower than the maximal-accuracy uncertainty; throughout the optimal uncertainty is less than the maximal-accuracy uncertainty}
    \label{fig:ulbd_bigger_uncertainty}
\end{figure}

These results are true in general, that the optimal uncertainty does not fall along the single threshold line.  We then further illustrate this for many different analyte to interferent ratios.  To do so, for each analyte to interferent ratio, a simplex based optimization method \cite{Troemel2018} was used to determine the location and value of the minimal uncertainty.  Then, the unbiased mean and its uncertainty band was plotted in \cref{fig:ulbd_uncertainty_size}, and the relative size of the optimal to the maximal-accuracy uncertainty band was plotted on \cref{fig:ulbd_bigger_uncertainty}.

These figures show that the optimal and maximal-accuracy activity estimates do not differ from each other, nor do they appreciably differ from the true value.  They show that at low analyte to interferent ratios, the optimal uncertainty is up to $3\times$ smaller than that for maximal accuracy; at high analyte to interferent ratios the uncertainty bands are close to each other, but still smaller in the optimal case.  Also of note is the size of the uncertainty bands for small analyte to interferent ratio.  At the lowest analyte to interferent ratios, the maximal-accuracy uncertainty band includes $0$ within two standard deviations. This echos the conclusions of \cref{sec:detection_limit}: that the maximal-accuracy thresholds do not have minimal detection limits. It also adds on stronger conclusions: in some cases, using maximal-accuracy thresholds precludes one from detecting what can be actually quantified using optimal thresholds.

\begin{figure*}
    \centering
    \includegraphics{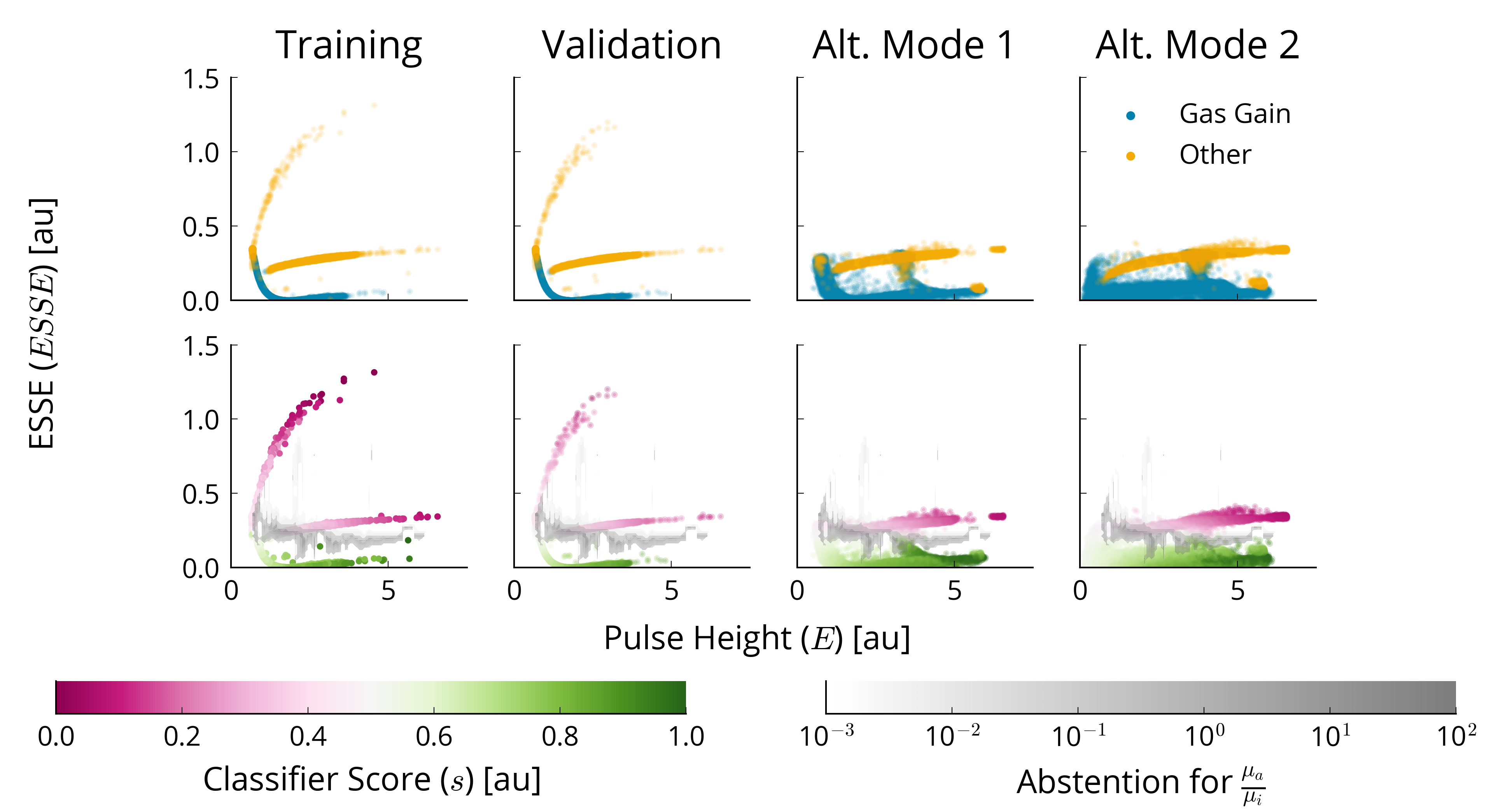}
    \caption{ESSE versus Pulse Height for different sets of detection events, and corresponding classifier scores for those events. The events along the boundary between the two classes have classifier scores closer to $\frac{1}{2}$, correspondingly abstention of these events happens at higher and higher analyte to interferent ratios.}
    \label{fig:ulbd_scatter}
\end{figure*}

\paragraph{Abstention on out of distribution data} As a final check, we are able to look at types of pulses from which we abstain.  To investigate this, we present \cref{fig:ulbd_scatter}. In it, we plot each pulse decomposed into a Pulse Height versus Exemplar Squared Sum of Errors (ESSE) representation, similar to the pulse shape discrimination plots in \cite{Mace2018}.  It can be seen that the regions of phase space indicative of Gas Gain versus Other in the training and validation sets are separated, with some overlap on the bottom left side of each chart. It can also be seen that pulses close to that intersection have classifier scores closer to $\frac{1}{2}$, and that the abstention region neatly splits those two regions.  For the alternate mode data sets, the Gas Gain region and Other region are not as well separated; however the abstention region removes many of those points which are in ambiguous regions of phase space.  This shows not only the utility of abstention for traditional in-distribution data classification, but its extension when coupled with ODIN to an out-of-distribution data set.

\section{Conclusions}

Citing the utility of using classifiers to correct for difficult-to-reduce backgrounds in counting experiments, we claim that a classifier in the cases presented here is better judged by detection limit and minimal uncertainty than mere accuracy. We demonstrated that these are \emph{not} the same. We presented a derivation of the detection limit and measurement uncertainty in such classifier-based counting experiments.

We also presented and showed the utility of abstaining from certain events in CBCEs.  For detection limit calculations, direct optimization of the classifier score threshold above which an event is classified as analyte can make large differences in the detection limit.  For activity measurements, using two thresholds, one above which all events will be considered analyte and the other below which all events will be considered interferent, can reduce the uncertainty by several times, dependent on the base rate of interferent.

Overall, this paper concludes that caution is warranted when judging a classifier to be used in CBCEs, and accuracy should not be the primary goal. We presented best practices for two metrics more relevant to metrology, but our concept of abstention applies to and would improve other metrics as well.

\section{Acknowledgement}

This research was funded by the National Nuclear Security Administration, Office of Defense Nuclear Nonproliferation Research and Development.

\bibliographystyle{ieeetr}
\bibliography{references.bib}

\begin{appendices}
\section{Details of labeling for and design of Tail Over Total Classifier}\label{sec:totc_labeling}

In order to determine the difference in shape of electrical pulses measured from a $\ce{Cs}_{2}\ce{Li}\ce{Y}\ce{C}_{6}$ (CLYC) detector when exposed to neutrons and photons, the detector was exposed to high-intensity $\ce{^{252}Cf}$ and $\ce{^{137}Cs}$, respectively.  This does not generate an environment with only neutrons or photons present for several reasons: $\ce{^{252}Cf}$, while mostly a neutron source, also produces photons, and neutrons and photons are present from naturally occurring radioactive material and cosmogenic sources.  Therefore, determining associating the ground truth particle origin with a given event is difficult or impossible.  To determine a set of labels for each event, identifying each as from neutrons or photons, we used an approximate process.

We used a Uniform Manifold Approximation and Projection (UMAP) dimensionality reduction technique \cite{McInnes2018} to reduce the pulses into two components.  It was noticed that the first component varied with respect to the curvature around the maximum magnitude of the pulse. We set a threshold on that component, all pulses above which were labeled as neutron and below which were labeled as photon.  Through tests with data sets including exposure to high intensity photon sources, we beleive that this labeling scheme is $>99\%$ accurate. However, as noted in the text, the physical accuracy of this labeling scheme is not important to this work - abstention and detection limit minimization is equally applicable to \emph{any} set of labels.

The classifier for Tail over Total Classification (TOTC) is a multilayer perceptron, of which we use \texttt{scikit-learn}'s implementation \cite{scikit-learn}. TOTC has 5 layers with 10 nodes in each, and uses ReLU activation.  It is trained by minimizing categorical cross entropy between predictions and the labels as described above using an Adam optimizer with weight decay of $10^{-2}$. These parameters were chosen empirically to optimize the area under the receiver operating characteristic curve.

\section{Derivation of measurement uncertainty in a counting based classifier experiment}\label{sec:derive_uncertainty}

We perform a measurement and obtain a number of counts $X$, generated independently by analyte ($X_{a}$) and interferent ($X_{i}$), of which these counts may be modeled as having means $\mu$, $\mu_{a}$ and $\mu_{i}$; and variances $\sigma^{2}$, $\sigma_{a}^{2}$ and $\sigma_{i}^{2}$, respectively.  We classify each count using some classifier, generating counts $Y_{a|a}$ of truly analyte counts classified as analyte, $Y_{a|i}$ truly analyte counts classified as interferent, $Y_{i|a}$ truly interferent counts classified as analyte, and $Y_{i|i}$ truly interferent counts classified as interferent. The probability that a count in $X$ is correctly classified as analyte is denoted as $p_{a}$, and the probability that it is correctly classified as interferent is denoted as $p_{i}$. We donote their incorrect counterparts as $q_{a}$ and $q_{i}$, respectively.  Note that $p_{a} + q_{a} \leq 1$ and $p_{i} + q_{i} \leq 1$, where the inequality holds only when there is non-zero abstention.

We write the expected value of analyte count $Y_a = Y_{a|a} + Y_{a|i}$ as
\begin{equation}
    \mathbb{E}\left(Y_{a}\right) = \mathbb{E}\left(X_{a}p_{a} + X_{i}q_{i}\right) = \mu_{a}p_{a} + \mu_{i}q_{i}
\end{equation}
and the equivalent for interferent as
\begin{equation}
    \mathbb{E}\left(Y_{i}\right) = \mathbb{E}\left(X_{a}q_{a} + X_{i}p_{i}\right) = \mu_{a}q_{a} + \mu_{i}p_{i}
\end{equation}

Then, the method of moments allows us to obtain an unbiased estimator by first setting the measured analyte count equal to the expected analyte count and solving the system of equations for $\mu_{a}$ (which we now denote $\hat{\mu}_{a}$ because of the estimation of the expectation value). This obtains
\begin{equation}
    \hat{\mu}_{a} = \frac{p_{i}Y_{a} - q_{i}Y_{i}}{p_{a}p_{i} - q_{a}q_{i}}
\end{equation}

We estimate the variance in $\hat{mu}_{a}$ using the conditional variance formula \cite{Ross2014}:
\begin{equation}
    \mathrm{Var}\left(\hat{\mu}_{a}\right)=\mathbb{E}\left[\mathrm{Var}\left(\hat{\mu}_{a}|X_{a},X_{i}\right)\right] + \mathrm{Var}\left(\mathbb{E}\left[\hat{\mu}_{a}|X_{a},X_{i}\right]\right)
\end{equation}

Several of the random variables are multinomial (i.e. $X_{a}$ is a sum of $Y_{a|a}$, $Y_{a|i}$ and any unclassified counts which are truly analyte $Y_{a|u}$).  We find that the mean and variance for $Y_{a|a}$ in that case is $X_{a}p_{a}$ and $X_{a}p_{a}\left(1-p_{a}\right)$, and its covariance with $Y_{a|i}$ is $-X_{a}p_{a}q_{a}$.  Performing similar substitutions for all variables, we successively obtain
\begin{multline}
    \mathbb{E}\left[\mathrm{Var}\left(\hat{\mu}_{a}|X_{a}X_{i}\right)\right] = \mathbb{E}\left[\frac{1}{\left(p_{a}p_{i} - q_{a}q_{i}\right)^{2}}\right. \\
    \left\{\mathrm{Var}\left(p_{i}Y_{a|a}-q_{i}Y_{a|i}|X_{a}X_{i}\right)\right.\\
    \left.\left.+ \mathrm{Var}\left(p_{i}Y_{i|a}-q_{i}Y_{i|i}|X_{a}X_{i}\right)\right\}\right]
\end{multline}
and
\begin{multline}
    \mathrm{Var}\left(p_{i}Y_{a|a}-q_{i}Y_{a|i}|X_{a}X_{i}\right) = X_{a}\left[p_{i}^{2}p_{a}\left(1 - p_{a}\right)\right. \\
    \left.+ q_{i}^{2}q_{a}\left(1-q_{a}\right) + 2p_{i}p_{a}q_{i}q_{a}\right]
\end{multline}
\begin{multline}
    \mathrm{Var}\left(p_{i}Y_{i|a}-q_{i}Y_{i|i}|X_{a}X_{i}\right) = X_{i}\left[p_{i}^{2}q_{i}\left(1 - q_{i}\right)\right. \\
    \left.+ q_{i}^{2}p_{i}\left(1-p_{i}\right) + 2p_{i}^{2}q_{i}^{2}\right]
\end{multline}
Finally, we see that
\begin{equation}
    \mathrm{Var}\left(\mathbb{E}\left[\hat{\mu}_{a}|X_{a}X_{i}\right]\right) = \mathrm{Var}\left(X_{a}\right) = \sigma_{a}^{2}
\end{equation}
Which results in the overall equation for uncertainty in \cref{eq:uncertainty_unbiased_mean}, reprinted below for convenience.
\begin{multline}
    \mathrm{Var}\left(\hat{\mu}_{a}\right) = \sigma_{a}^{2} + \frac{1}{\left(p_{a}p_{i} - q_{a}q_{i}\right)^{2}} \\
    \cdot \left\{\hat{mu}_{a}\left[p_{i}^{2}p_{a}\left(1-p_{a}\right) + q_{i}^{2}q_{a}\left(1-q_{a}\right) + 2p_{i}p_{a}q_{i}q_{a}\right]\right. \\
    \left. + \mu_{i}\left[p_{i}^{2}q_{i}\left(1-q_{i}\right)+q_{i}^{2}p_{i}\left(1-p_{i}\right) + 2p_{i}^{2}q_{i}^{2}\right]\right\}
\end{multline}
In the text, we use $\mu_{a}$ and $\mu_{i}$ as notation for the estimated means (instead of $\hat{\mu}_{a}$ and $\hat{\mu}_{i}$), as we never reference the true mean in the text.

\section{Details of ultra-low background proportional counter classifier}\label{sec:ulbpc}

To create the classifier separating Gas-Gain from Other events in the ultra-low background proportional counter data set, we create a multilayer perceptron of 5 layers, with 100 nodes per layer and activated with ReLU, following best practices as laid out in \cite{Goodfellow-et-al-2016} and other sources. We use the \texttt{pytorch} framework \cite{pytorch} and implement a Generalized ODIN architecture \cite{Hsu2020} to encourage low scores for data seen during inference which appears to be out-of-distribution from the training set. This method requires that the penultimate layer of the neural network is split explicitly into a numerator and denominator; their quotient is then minimized by cross-entropy against the set of labels.  The explicit separation of numerator and denominator follows the law of conditional probability to decompose the model output into the conditional probability that a pulse is both a Gas Gain pulse and is from in-distribution data (the numerator); and the probability that the pulse is from in-distribution data (the denominator).  Then, we can use the numerator as our classifier score. The score be close to $\frac{1}{2}$ if the classifier is unsure about which class it belongs in, or if it comes from out-of-distribution data.

\end{appendices}

\end{document}